\documentclass[letterpaper, 10 pt, conference]{ieeeconf}  

\IEEEoverridecommandlockouts                              
\usepackage{cite}
\let\proof\relax 
\let\endproof\relax
\usepackage{amsmath,amssymb,amsfonts}
\usepackage{graphicx}
\usepackage{textcomp}
\usepackage{xcolor}
\usepackage{amsmath, amssymb, amsfonts}  
\usepackage{algorithm}
\usepackage{algpseudocode}
\usepackage{tikz}
\usetikzlibrary{positioning, arrows, fit}
\usepackage{booktabs} 
\usepackage{multirow}
\usepackage{url}
\usepackage{amsthm}

\usepackage{mathtools}
\usepackage{dsfont}
\usepackage{comment}
\usepackage{flushend}

\newboolean{arxivversion}

\setboolean{arxivversion}{true}  

\usepackage{setspace}

\DeclareMathAlphabet{\pazocal}{OMS}{zplm}{m}{n}
\newtheorem{assumption}{Assumption} 
\newtheorem{lemma}{Lemma}
\newtheorem{theorem}{Theorem}
\newtheorem{proposition}{Proposition}

\newtheorem{remark}{Remark}

\title{\LARGE \bf
Robust MPC for Uncertain Linear Systems -\\ Combining Model Adaptation and Iterative Learning 
} 

\author{Hannes Petrenz$^1$, Johannes Köhler$^2$, and Francesco Borrelli$^3$
\thanks{$^{1}$ Student at the University of Stuttgart, 70174 Stuttgart, Germany, Visiting Student Researcher at the Department of Mechanical Engineering, University of California at Berkeley, Berkeley, CA 94701, USA
         {\tt\small st161475@stud.uni-stuttgart.de}}
\thanks{$^{2}$ Institute for Dynamic Systems and Control, ETH Zurich, 8052 Zürich, Switzerland,
         {\tt\small jkoehle@ethz.ch}}
\thanks{$^{3}$ Department of Mechanical Engineering, University of California at Berkeley , Berkeley, CA 94701, {\tt\small fborrelli@berkeley.edu}}}

\begin{document}

\maketitle
\thispagestyle{empty}
\pagestyle{empty}

\begin{abstract}
This paper presents a robust adaptive learning Model Predictive Control (MPC) framework for linear systems with parametric uncertainties and additive disturbances performing iterative tasks. The approach refines the parameter estimates online using set-membership estimation. Performance enhancement over iterations is achieved by learning the terminal cost from data. Safety is enforced using a terminal set, which is also learned iteratively. The proposed method guarantees recursive feasibility, constraint satisfaction, and a robust bound on the closed-loop cost. Numerical simulations on a mass-spring-damper system demonstrate improved computational efficiency and control performance compared to a robust adaptive MPC scheme without iterative learning of the terminal ingredients.
\end{abstract}

\section{INTRODUCTION}
Model predictive control (MPC)~\cite{rawlings2017model} is an established optimization-based control technique, which is widely used for systems subject to input and state constraints.
When dealing with iterative tasks, data from previous iterations can be leveraged to estimate model parameters~\cite{tanaskovic2014adaptive} and enhance the performance~\cite{rosolia2019learning} of the MPC scheme.
This paper investigates an MPC formulation that guarantees robust constraint satisfaction while simultaneously improving performance by adapting to unknown parameters and learning from previous iterations.

Robust MPC schemes~\cite{saltik2018outlook, kouvaritakis2016model}  ensure robust constraint satisfaction for bounded model mismatch.
Tube-based robust MPC formulations~\cite{langson2004robust}
offer these robustness guarantees with a particularly favorable computational complexity.
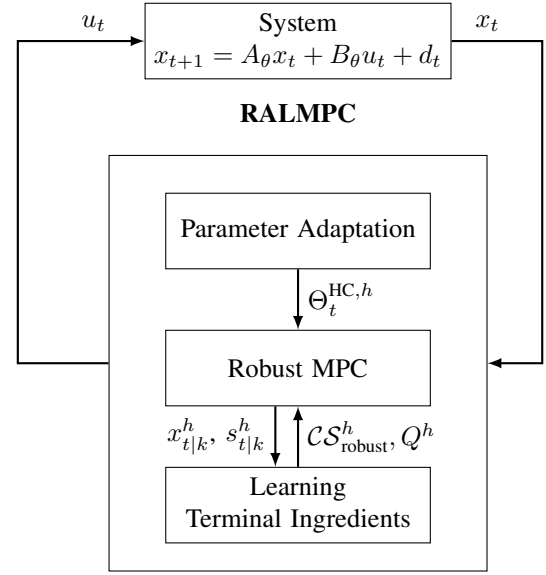
\begin{figure}[ht]
    \centering
    \begin{tikzpicture}[
        block/.style={draw, rectangle, minimum width=4cm, minimum height=1cm, align=center},
        arrow/.style={-latex, thick}
    ]

    \node (system) [block] {System \\ $x_{t+1} = A_\theta x_t + B_\theta u_t + d_t$};

    \node (ralmpc) [draw, rectangle, below=1.0cm of system, minimum width=5cm, minimum height=5.5cm, anchor=north] {};

    \node at (ralmpc.north) [above=0.3cm] {\textbf{RALMPC}};  

    \node (identification) [block, below=0.5cm of ralmpc.north, minimum width=3.5cm] 
    {Parameter Adaptation};

    \node (mpc) [block, below=0.8cm of identification, minimum width=3.5cm] 
    {Robust MPC};

    \node (learning) [block, below=0.8cm of mpc, minimum width=3.5cm] 
    {Learning \\ Terminal Ingredients};

    \node (lower) [anchor=north, below=0.8cm of learning] {};

    \draw [arrow] (system.east) -- ++(1.2,0) node[pos=0.4, above] {$x_t$} |- (ralmpc.east);
    \draw [arrow] (ralmpc.west) -- ++(-1.2,0) |- (system.west) node[pos=0.8, above] {$u_t$};

    \draw [arrow] (identification.south) -- node[right] {$\Theta^{\text{HC},h}_t$} (mpc.north);

    \draw [arrow] (learning.north) -- node[midway, right] {$\mathcal{CS}^h_{\text{robust}}, Q^h$} (mpc.south);
    \draw [arrow] ([xshift=-0.3cm] mpc.south) -- ([xshift=-0.3cm] learning.north) 
        node[midway, left] {$x_{t|k}^h$, $s_{t|k}^h$};
    \end{tikzpicture}
     \vspace{-0.7cm}
    \caption{Schematic of the proposed robust adaptive learning Model Predictive Control (RALMPC) framework: At time $t$, the RALMPC controller receives the state $x_t$ and computes the control input $u_t$. The parameter set $\Theta^{\text{HC},h}_t$ is then updated via set-membership estimation using $x_t$. Based on the predicted states $x_{t|k}^h$ and $s_{t|k}^h$, the framework constructs the terminal set $\mathcal{CS}^h_{\text{robust}}$ and the terminal cost function $Q^h$.}
    \label{fig:ralmpc_diagram}
    \vspace{-0.7cm}
\end{figure}
These approaches achieve robust constraint satisfaction by 
 enclosing all possible trajectories within a tube around the nominal trajectory using a local feedback law. 
 
However, while tube-based robust MPC schemes perform well under disturbances, they can become overly conservative in the presence of constant parametric uncertainties. This limitation has increased interest in the development of robust adaptive MPC, in which uncertain parameters are adapted online using past data~\cite{kim2008adaptive}.
In~\cite{aswani2013provably}, safety and performance are decoupled via two separate models, with only the performance model adapted online.
Set-membership estimation techniques have been employed to update models while ensuring recursive feasibility for time-invariant, time-variant, and stochastic finite impulse response systems~\cite{tanaskovic2014adaptive,tanaskovic2019adaptive,bujarbaruah2018adaptive}. However, these methods apply only to a limited class of asymptotically stable systems. This limitation motivated the authors of~\cite{lorenzen2019robust} to extend these results to a broader class of linear state-space systems. Nevertheless, the resulting robust adaptive MPC formulation requires optimization over an increasing number of variables and constraints. To address this,~\cite{kohler2019linear} and~\cite{lu2019robust} develop robust adaptive MPC algorithms with fixed computational complexity during runtime. 
Furthermore,~\cite{didier2021robust} demonstrates an experimental application of~\cite{kohler2019linear} to a quadrotor system.

In many applications, the same control task is solved repeatedly, making it advantageous to leverage information from previous iterations in the controller design. Iterative Learning Control~\cite{bristow2006survey} addresses this by refining the controller in each iteration.~\cite{rosolia2017learning,rosolia2017learning_2,rosolia2019learning} develop an iterative learning MPC scheme that utilizes past experiments to define a terminal cost and terminal set. By incorporating these terminal ingredients into the MPC formulation, convergence to an improved closed-loop cost function is shown. 
The recent work~\cite{hashimoto2024robust} combines iterative learning with system identification for Lipschitz-continuous nonlinear systems, though it has a large computational demand. 
A modular framework for learning-based control of uncertain linear systems is proposed in~\cite{didier2021adaptive}, ensuring robust safety by iteratively refining a safe set using robust tubes and improved model estimates. 
In~\cite{bujarbaruah2018adaptive}, set-membership estimation is combined with iterative learning for linear systems. However, this method is limited to constant parametric offsets and cannot treat general parametric uncertainty and disturbances.

In this paper, we propose a new robust adaptive learning MPC (RALMPC) algorithm that combines the robust adaptive MPC framework of~\cite{kohler2019linear} with the iterative learning approach of~\cite{rosolia2017learning_2}. The core idea is to adaptively estimate uncertain parameters online using set-membership estimation, while simultaneously refining the control policy for iterative tasks using data from previous iterations. The algorithm is designed for linear state-space systems subject to constant parametric uncertainties and additive bounded disturbances. Figure~\ref{fig:ralmpc_diagram} provides a schematic overview of the different elements of the algorithm.

The key contributions of this work are as follows:
\begin{itemize}
    \item A robust adaptive learning MPC algorithm is developed for linear time-invariant systems subject to parametric uncertainties and additive disturbances performing iterative tasks. 
    \item A robust control invariant set is constructed using previously predicted trajectories. The set is iteratively expanded by adding new trajectories. Additionally, a terminal cost based on the worst-case cost-to-go of the predicted trajectories is constructed. This approach enables the controller to (robustly) improve the closed-loop cost over iterations. 
    \item The proposed approach ensures recursive feasibility and robust constraint satisfaction. In addition, a robust bound on the closed-loop cost is established.
    \item A numerical example illustrates the effectiveness of the proposed algorithm, comparing it with the approach in~\cite{kohler2019linear}. The results highlight the advantages of a shorter prediction horizon, leading to reduced computational burden, as well as improved closed-loop cost due to the improvement of terminal conditions over iterations.
\end{itemize}
The remainder of this paper is structured as follows. Section~\ref{Problem Statement} presents the problem formulation. Section~\ref{RALMPC} introduces the parameter learning, constraint tightening, and terminal conditions. Section~\ref{Theoretical Analysis} provides a theoretical analysis of the resulting MPC scheme. Section~\ref{Simulation Results} presents numerical simulations, followed by concluding remarks in Section~\ref{Conclusion}.

\textit{Notation:} 
We denote $[A]_i$ as the $i$-th row of matrix $A$. The quadratic norm with the positive definite matrix $Q \succ 0$ is defined as $\|x\|_Q^2 = x^\top Q x$. The unit hypercube is denoted by $\mathbb{B}_p = \{\theta \in \mathbb{R}^p |~ \lVert \theta \rVert_{\infty} \leq 1 \}$ and the Minkowski sum is denoted by $A \oplus B$. The cardinality of set $A$ is denoted by $|A|$. $\mathcal{K}_\infty$ is a class of functions $\alpha$: $\mathbb{R}_{\geq0}\rightarrow\mathbb{R}_{\geq0}$, which are continuous, strictly increasing, unbounded and satisfy $\alpha(0)=0$.

\section{PROBLEM STATEMENT} \label{Problem Statement}
Consider the discrete-time uncertain linear system
\begin{equation}
    \label{eq: linear system}
    x_{t+1} = A_{\theta} x_t + B_{\theta} u_t + d_t, \quad x_0 = x_s,   
\end{equation}
where $x_t \in \mathbb{R}^n$ and $u_t \in \mathbb{R}^m$ are state and input at time $t\in\mathbb{N}$. The system is affected by additive disturbances $d_t \in \mathbb{R}^n$ and an unknown constant parameter $\theta=\theta^* \in \mathbb{R}^p$.
We impose mixed state and input constraints of the form
\begin{equation}
    \label{eq: constraints}
    (x_t, u_t) \in  \mathcal{Z} \quad \forall t\in\mathbb{N},
\end{equation}
where
$\mathcal{Z}= \{(x,u) \in \mathbb{R}^{n+m} \mid F_j x + G_j u \leq 1, \, j = 1, \dots, q\}$ is a compact set. 
Moreover, the uncertainties and disturbances satisfy the following assumption.
\begin{assumption}
\label{ass: general} 
\
\begin{itemize}
    \item The disturbances $d_t$ lie in a compact, polytopic set $\mathbb{D}=\{d \in \mathbb{R}^n \mid H_d d \leq h_d \}$ for all time $t\in\mathbb{N}$.
    \item The uncertain parameters $\theta \in \mathbb{R}^p$ enter affinely into the system matrices:
    \begin{equation}
    A_\theta = A_0 + \sum_{i=1}^p [\theta]_i A_i, \quad B_\theta = B_0 + \sum_{i=1}^p [\theta]_i B_i.
    \end{equation}
    \item The parameters $\theta$ are bounded in a prior known convex hypercube set $\Theta_0^{\mathrm{HC}} \in \bar{\theta}_0 \oplus \eta_0 \mathbb{B}_p$ with scaling factor $\eta_0\geq 0$ and center point $\bar{\theta}_0$. 
\end{itemize}
\end{assumption}

Throughout this paper, we focus on robustly stabilizing the origin using a quadratic stage cost $\ell(x, u) = \|x\|_Q^2 + \|u\|_R^2$, where $Q$ and $R$ are positive definite.
We consider an iterative scenario in which each iteration, denoted by \( h \in \mathbb{N} \), starts from the same initial state $x_s$. 
Our goal is to design a controller that improves control performance over iterations while robustly ensuring safety.
Therefore, we consider the following conceptual infinite horizon control problem:  
\begin{align}
\label{eq: problem}
&\min_{u^h_{\cdot}} \sum_{t=0}^\infty \max_{x \in \mathbb{X}_t^h} \ell(x, u_t^h(x))\\
 &\text{subject to:} \nonumber\\
 & \mathbb{X}_0^h \ni x_0^h = x_s, \nonumber\\
& \mathbb{X}_{t+1}^h \ni A_\theta x + B_\theta u_t^h(x) + d, \, \forall x \in \mathbb{X}_t^h, \; d \in \mathbb{D}, \; \theta \in \Theta_t^{\text{HC}, h}, \nonumber\\
& (x, u_t^h(x)) \in \mathcal{Z} \quad \forall x \in \mathbb{X}_t^h, \quad t \in \mathbb{N}, \nonumber\\
& 
\end{align}
where $\mathbb{X} ^h_{t}$ is a tube that outer bounds all possible state trajectories, $\Theta^{\text{HC},h} \subseteq \Theta_0^{\text{HC}}$ is the hypercubic parameter set at the $h$-iteration, and $u^h_{t}(x)$ is some feedback policy.   Since problem \eqref{eq: problem} is not tractable, we present a robust adaptive learning MPC scheme that approximates the optimal solution to this problem.
\section{Robust Adaptive Learning MPC} \label{RALMPC}
In Section~\ref{Set membership estimation}, we introduce the framework for parameter adaptation. This is followed by the introduction of the polytopic tube based on the previous work of~\cite{kohler2019linear} in Section~\ref{Polytopic tube} and the worst-case stage cost in Section~\ref{Worst case stage cost}. The robust adaptive learning MPC algorithm is presented in Section~\ref{RALMPC scheme}. Section~\ref{Terminal conditions} discusses the learning terminal conditions, which are the main theoretical contribution of this paper. Finally, Section~\ref{Algorithms} presents the offline and online computation algorithms.

\subsection{Set-membership estimation}\label{Set membership estimation}
Set-membership estimation reduces the feasible parameter set $\theta^*\in \Theta_t^{\text{HC}}\subseteq\Theta_0^{\text{HC}}$ by using past data. Since the controller must satisfy the constraints for all $\theta \in \Theta_t^{\text{HC}}$, reducing the size of $\Theta_t^{\text{HC}}$ leads to a decrease in conservatism, indicated by reduced tube size. 
We define the matrix
\begin{equation}
\label{eq:D_xu}
    D(x, u) := [A_1 x + B_1 u, \dots, A_p x + B_p u] \in \mathbb{R}^{n \times p},
\end{equation}
and the non-falsified parameter set
\begin{equation}
    \Delta_t := \left\{ \theta \in \mathbb{R}^p \,\middle|\, x_{t} - A_\theta x_{t-1} - B_\theta u_{t-1} \in \mathbb{D} \right\},
\end{equation}
which is the polytopic set containing all parameters consistent with the data point $(x_{t-1},u_{t-1},x_{t})$.
In a moving-horizon fashion, Algorithm~\ref{alg:hypercube} from~\cite{kohler2019linear} uses the past $M\in\mathbb{N}$ non-falsified parameter sets along with $\Theta_{t-1}^{\text{HC}}$ to compute a smaller over-approximating hypercube $\Theta_{t}^{\text{HC}}$.
\begin{algorithm}
\caption{Moving window hypercube update~\cite[Alg. 1]{kohler2019linear}}
\label{alg:hypercube}
Input: $\{\Delta_k\}_{k=t,\dots,t-M-1},\, \Theta_{t-1}^{\text{HC}}.$ Output: $\Theta_t^{\text{HC}}=\bar{\theta}_t \oplus \eta_t \mathbb{B}_p$
\begin{algorithmic}
\Statex Define polytope $\Theta_t^M := \Theta_{t-1}^{\text{HC}} \bigcap_{k=t-M-1}^{t} \Delta_k$
\Statex Solve $2p$ LPs ($i=1,\cdots,p$):
\Statex \, $\theta_{i,t,\min} := \min_{\theta \in \Theta_t^M} e_i^\top \theta$, $\theta_{i,t,\max} := \max_{\theta \in \Theta_t^M} e_i^\top \theta$
\Statex \, with unit vector $e_i = [0, \dots, 1, \dots, 0] \in \mathbb{R}^p, \quad [e_i]_i = 1.$
\Statex Set $[\bar{\theta}_t]_i = 0.5(\theta_{i,t,\min} + \theta_{i,t,\max})$
\Statex Set $\eta_t = 0.5 \max_i (\theta_{i,t,\max} - \theta_{i,t,\min})$
\Statex Project: $\bar{\theta}_t$ on $\bar{\theta}_{t-1} \oplus (\eta_{t-1} - \eta_t) \mathbb{B}_p$
\end{algorithmic}
\end{algorithm}
In contrast to the update rule $\Theta_t := \Theta_{t-1}\cap \Delta_t$, which can lead to an unbounded number of (non-redundant) half-spaces, Algorithm~\ref{alg:hypercube} provides a fixed complexity description of $\Theta_{t}^{\text{HC}}$ using a hypercube.
Moreover, this fixed parameterization ensures that the parameter update satisfies
\begin{equation}
\label{eq: bound on parameter update}
 \bar{\theta}_t - \bar{\theta}_{t-1} \in (\eta_{t-1} - \eta_t) \mathbb{B}_p,   
\end{equation}
where $\eta_{t-1} - \eta_t\geq0$. This is one of the key properties used in the analysis in Section~\ref{Theoretical Analysis}.
In addition, it enables us to state the following lemma:
\begin{lemma}
    \label{lem: Theta}
    Let Assumption~\ref{ass: general} hold and consider Algorithm~\ref{alg:hypercube}. The recursively updated sets $\Theta_t^{\text{HC}}$ contain the true parameters $\theta^*$ and satisfy $ \Theta_t^{\text{HC}} \subseteq \Theta_{t-1}^{\text{HC}}$ $\forall t\in\mathbb{N}$.
\end{lemma}
More details on Algorithm~\ref{alg:hypercube} and the proof of Lemma~\ref{lem: Theta} can be found in~\cite[Sec.~2.B]{kohler2019linear}

\subsection{Polytopic tube}\label{Polytopic tube}
Tube-based MPC bounds the state $x$ for all possible realizations of $d_t\in \mathbb{D}$ and $\theta \in \Theta_t^{\text{HC}}$ within a tube $\mathbb{X}$ around the nominal predicted trajectory. Thus, the approach captures the propagation of uncertainty and is crucial for ensuring safety. To reduce online computation, a fixed, offline-computed polytope 
\begin{equation}
    \label{eq: polytope}
     \mathcal{P} = \left\{x \in \mathbb{R}^n \mid H_i x \leq 1, \; i = 1, \dots, r \right\},
\end{equation}
is used to parametrize the tube $\mathbb{X}$. In particular, the tube is obtained by scaling this fixed polytope online by a scalar dilation $s \geq 0$ and centering it at a nominal prediction $z$, i.e. $\mathbb{X}=z\oplus s \mathcal{P}$.\\
To reduce conservatism, we employ a pre-stabilizing feedback
$Kx$ that satisfies the following assumption.
\begin{assumption}
\label{ass: controller}
\textit{There exist a feedback matrix $K \in \mathbb{R}^{m \times n}$ and a positive definite matrix $P$, such that $A_{\text{cl},\theta} := A_{\theta} + B_{\theta}K$ is quadratically stable~\cite[Def. 1]{acikmese2003stability} and satisfies}
\[
A_{\text{cl},\theta}^{\top} P A_{\text{cl},\theta} + Q + K^{\top} R K \preceq P, \quad \forall \theta \in \Theta_0^{\text{HC}},
\]
\textit{with the prior parameter set $\Theta_0^{\text{HC}}$ from Assumption 1.}
\end{assumption}
The technical properties of tube propagation can be found in~\cite{kohler2019linear} and are summarized in the following lemma.
\begin{lemma}
\label{lem: Tube propagation}
    Let Assumptions~\ref{ass: general} and~\ref{ass: controller} hold.
    Define the following constants:
    \begin{align}
        \rho_{\bar{\theta}} &\coloneq \underset{i}{\max} \ \underset{x \in \mathcal{P}}{\max} H_i A_{\text{cl},\bar{\theta}} x, \label{eq:contraction rate}\\ 
        L_{\mathbb{B}} &\coloneq \underset{i,l}{\max} \ \underset{x \in \mathcal{P}}{\max} H_i D(x, Kx) \tilde{e}_l, \label{eq: lipschitz}\\ 
        \bar{d} &\coloneq \underset{i}{\max} \ \underset{d \in \mathbb{D}}{\max} H_i d ,\label{eq: disturbance bound}
    \end{align}
    where $\tilde{e}_l\in\mathbb{R}^p$ denotes the $2^p$ vertices of the unit hypercube $\mathbb{B}_p$.
    Recalling that the effect of the model parameters $\theta$ on the dynamics is given by $D(x,u)(\theta^*-\bar{\theta})$ with $D(x,u)$ according to~\eqref{eq:D_xu}, we define the function:
    \begin{align}
        w_{\eta}(z,v) \coloneq \eta \underset{i,l}{\max} H_i D(z, v) \tilde{e}_l.\label{eq: parametric uncertainty bound}
    \end{align}
    Then, for any \( z \in \mathbb{R}^n \) , \( v \in \mathbb{R}^m\), and \( \Theta^{\text{HC}} = \bar{\theta} \oplus \eta \mathbb{B}_p \) with  \( \bar{\theta} \in \mathbb{R}^p \),\( \eta \geq 0 \), and for any \( x \in z \oplus s \mathcal{P} \) with \( s \geq 0 \), it holds that
    $
        x^+ \in z^+ \oplus s^+ \mathcal{P} \label{eq: tube step plus}
    $
    with:
    \begin{align}
        z^+ &= A_{\text{cl},\bar{\theta}}z + B_{\bar{\theta}}v, \label{eq: nominal dynamic}\\ 
        s^+ &=(\rho_{\bar{\theta}} + \eta L_{\mathbb{B}}) s + \bar{d} + w_{\eta}(z,v), \label{eq: tube dynamic}\\ 
        x^+ &= A_{\text{cl},\theta} x + B_{\theta} v + d, \quad \forall \theta \in \Theta^{\text{HC}}, \quad d \in \mathbb{D}\label{eq: uncertain dynamic}. 
    \end{align}
\end{lemma}
This lemma states that the state $x^+$ described by the dynamic~\eqref{eq: uncertain dynamic} is contained in the tube $s^+\mathcal{P}$ around the nominal state $z^+$, which evolves according to the dynamics~\eqref{eq: nominal dynamic}, where $s^+$ follows the scalar tube dynamics~\eqref{eq: tube dynamic}.
The following assumption ensures that the tube propagation in Lemma~\ref{lem: Tube propagation} yields a bounded scaling $s$, which is related to the choice of polytope $\mathcal{P}$.
\begin{assumption}
\label{ass: contractive}
The polytope $\mathcal{P}$ is chosen such that
\begin{equation}
    1 > \rho_{\bar{\theta}_0} + \eta_0 L_{\mathbb{B}}.
\end{equation}
\end{assumption}

Finally, recalling the mixed constraints $F_j x + G_j u \leq 1$, we define the following constants, which will be useful later in this paper:
\begin{equation}
\label{eq: constraint bound}
    c_j:= \underset{x \in \mathcal{P}}{\max} [F+GK]_j x, \quad  j=1, \dots, q.
\end{equation} 

\subsection{Worst-case stage cost} \label{Worst case stage cost}
We define the worst-case stage cost as 
\begin{equation}
    \label{eq: worst case stage cost}
    \ell_{\mathrm{max}}(x, v, s) := \ell(x, Kx + v) + L_{\text{cost}} s, 
\end{equation}
where $L_{\text{cost}} \geq 0$ is chosen such that for all $(x, Kx + v) \in \mathcal{Z}$ and $(z, Kz + v) \in \mathcal{Z}$, with $s = \max_i H_i(x - z)$, the following holds:
\begin{equation}
 \label{eq: Lipschitz constant cost}
  \ell(z,Kz+v)\leq\ell(x,Kx+v)+L_{\text{cost}}s.
\end{equation}
Since the quadratic cost $\ell$ is Lipschitz continuous on the compact set $\mathcal{Z}$, such a constant $L_{\text{cost}}$ exists and can be computed analogously to a Lipschitz constant.
This cost satisfies the following monotonicity property:
\begin{equation}
    \ell_{\text{max}}(x,v,s)\leq  \ell_{\text{max}}(\tilde{x},v,\tilde{s}), \label{eq: decrease stage cost}
\end{equation}
for any $x$, $s$, $\tilde{x}$, $\tilde{s}$ satisfying $x \oplus s \mathcal{P} \subseteq  \tilde{x} \oplus \tilde{s} \mathcal{P}$.
Furthermore, we define the steady-state cost for $x_{\text{steady}}=0$ and $v_{\text{steady}}=0$ as
\begin{equation}
    \label{eq: worst case steady state stage cost}
    \ell_{\text{max,s}}=L_{\text{cost}}s_{\text{steady}}, 
\end{equation}
with $s_{\text{steady}}=\bar{d}/(1-(\rho_{\bar{\theta}_0}+\eta_0 L_{\mathbb{B}}))$.

\subsection{Robust adaptive learning MPC scheme}\label{RALMPC scheme}
 At each time step $t$ of the $h$-iteration, given the state $x_t^h$, the current hypercube $\Theta_t^{\text{HC},h} = \bar{\theta}_t^h \oplus \eta_t^h \mathbb{B}_p$, the constants $\rho_{\bar{\theta}_t^h}$, $L_{\mathbb{B}}$, $\bar{d}$, $K$, and the function $w_{\eta_t^h}(z,v)$, the proposed robust adaptive learning MPC scheme is formulated as follows:
\begin{subequations}
\label{eq:RALMPC}
\begin{align}
 &\min_{v_{\cdot|t}^h,\lambda_{t}^h} \sum_{k=0}^{N-1} \ell_{\text{max}}(x_{k|t}^h,v_{k|t}^h,s_{k|t}^h) + Q^{h-1}(\lambda_{t}^h)  \notag\\  
&\text{subject to} \notag\\
&  \ x_{0|t}^h = x_t^h, \quad s_{0|t}^h = 0,  \label{eq:RALMPC a}\\
& \ x_{k+1|t}^h = A_{\text{cl},\bar{\theta}_{t}^h} x_{k|t}^h + B_{\bar{\theta}_{t}^h} v_{k|t}^h, \label{eq:RALMPC b}\\
& \  s_{k+1|t}^h = \rho_{\bar{\theta}_{t}^h} s_{k|t}^h + w_{k|t}^h, \label{eq:RALMPC d}\\
& \  w_{k|t}^h=\bar{d}+\eta_{t}^hL_{\mathbb{B}}s_{k|t}^h+w_{\eta_{t}^h}(x_{k|t}^h,u_{k|t}^h), \label{eq:RALMPC e}\\
& \  F_j x_{k|t}^h + G_j u_{k|t}^h + c_j s_{k|t}^h \leq 1, \label{eq:RALMPC f} \\   
& \  u_{k|t}^h = v_{k|t}^h + Kx_{k|t}^h, \label{eq:RALMPC g}\\   
& \  (x_{N|t}^h, s_{N|t}^h,\lambda_{t}^h) \in \mathcal{CS}_{\text{robust}}^{h-1}, \label{eq:RALMPC h}\\    
& \ k=0,\dots,N-1,\quad  j=1,\dots,q.
\end{align}
\end{subequations}
The scalar dynamics in Equations~\eqref{eq:RALMPC d}--\eqref{eq:RALMPC e} describe the propagation of the tube from Lemma~\ref{lem: Tube propagation}.
Based on $x^h_{k|t}$, $s^h_{k|t}$, and $v_{k|t}^h$, the cost function minimizes an upper bound on the stage cost. 
The tightened constraints~\eqref{eq:RALMPC f} with $c_j$ in~\eqref{eq: constraint bound} ensure that all trajectories in $\mathbb{X}_{k|t}$ satisfy the constraints. 

The terminal cost $Q^{h-1}(\lambda_{t}^h) $ and the terminal set $\mathcal{CS}_{\text{robust}}^{h-1}$ are constructed at each iteration from data collected in previous iterations, as specified in the following section. The optimal solution is indicated by $v_{k|t}^{h,*}$, $w_{k|t}^{h,*}$, $\lambda_{t}^{h,*}$, $x_{k|t}^{h,*}$, $ s_{k|t}^{h,*}$, $u_{k|t}^{h,*}$ with corresponding optimal cost function $J^{\text{LMPC},h,*}(x_t^h,\Theta_t^{\text{HC},h})$. The closed-loop input is given by $u_t^h=Kx_t^h+v_{0|t}^{h,*}$.

\subsection{Terminal conditions}\label{Terminal conditions}
The terminal condition~\eqref{eq:RALMPC h} leverages
 past prediction trajectories to mitigate the limitation of the finite prediction horizon $N$.
To construct this set, we adopt the following assumption.
\begin{assumption}
\label{ass:initial_solution}
   We have access to a finite-horizon trajectory \([x_s, x_1^0, \dots, x_{\bar{N}}^0]\), \([v_0^0, \dots, v_{\bar{N}-1}^0]\), and \([0, s_1^0, \dots, s_{\bar{N}}^0]\), which satisfies~\eqref{eq:RALMPC b}--\eqref{eq:RALMPC f} for \(k = 0, \dots, \bar{N}-1\), and
\begin{equation}
x_{\bar{N}}^0 = x_{\text{steady}} = 0,\quad s_{\bar{N}}^0 = s_{\text{steady}} = \frac{1}{1 - (\rho_{\bar{\theta}_0} + \eta_0 L_{\mathbb{B}})} \bar{d},
\end{equation}
with \( x_k^0 = 0 \), \( s_k^0 = s_{\text{steady}} \), and \( v_k^0 = 0 \) for \( k \geq \bar{N} \).
\end{assumption}
The terminal constraint \( \mathcal{CS}^h_{\text{robust}} \) is constructed from data to serve as a robust control-invariant set.  
By adding \( (x_{k|t}^h, s_{k|t}^h) \) for all \( k = 0, \dots, N-1 \) at time \( t \) of the \( h \)-iteration to the set, we enlarge it at runtime, thereby expanding the solution space and enabling learning. In the iterative learning MPC~\cite{rosolia2017learning_2}, a terminal set is constructed with measured closed-loop trajectories, assuming no model mismatch. Instead, we utilize the tube prediction of~\eqref{eq:RALMPC b} and~\eqref{eq:RALMPC d}. 
We define the sample set at $h$-iteration as
\begin{equation}
\label{eq: sample set}
    \mathcal{SS}^h = \left\{ \bigcup_{\hat{h}=0}^h \bigcup_{\hat{t}=0}^{\infty} \bigcup_{\hat{k}=0}^{N-1} (x_{\hat{k}|\hat{t}}^{\hat{h}}, s_{\hat{k}|\hat{t}}^{\hat{h}}) \right\},
\end{equation}
and the convex sample set
\begin{align}
    \mathcal{CS}^h=& \left\{  (z,\lambda)\in\mathbb{R}^{n+1+|\mathcal{SS}^h|} \;\middle|\; \right.\\
    &\quad \left. z=\sum_{i=1}^{|\mathcal{SS}^h|} \lambda_iz_i, \lambda_i\geq0,\sum_{i=1}^{|\mathcal{SS}^h|} \lambda_i=1, z_i \in \mathcal{SS}^h \ \right\},\notag
\end{align}
where Assumption~\ref{ass:initial_solution} ensures that both sets are non-empty at $h=0$.
The robust convex safe set 
\begin{align}
     \mathcal{CS}^h_{\text{robust}} =& \left\{ (x, s, \lambda) \in \mathbb{R}^{n+1+|\mathcal{SS}^h|}\;\middle|\;  \exists \; (x', s',\lambda) \in \mathcal{CS}^h\; \text{s.t.} \right. \notag\\ 
     &\quad \left. \underset{i}{\max} H_i(x - x') \leq s'-s \;  \right\},\label{eq: convex sample safe set}
\end{align}
is constructed such that $\mathbb{X} = \{z \mid H_i (z - x)  \leq s\} \subseteq \mathbb{X} '= \{z \mid H_i (z - x')  \leq s'\}$. 
Given the initial trajectory, we define the terminal cost for the first iteration as
\begin{equation}
\label{eq:initial terminal cost}
\begin{split}
    Q^{0}(\lambda)&= \sum_{\hat{k}=0}^{\infty} \lambda_{\hat{k}|0}^{0} \mathcal{J}^{0}_{\text{wc},\hat{k}|0},\\
    \text{with}& \quad\mathcal{J}^{0}_{\text{wc},\hat{k}|0}=\sum_{k=\hat{k}}^{\bar{N}-1}\ell_{\text{max}}(x^{0}_{k},v^{0}_{k},s^{0}_{k})-\ell_{\text{max,s}},\\
    &\quad\lambda=[\dots,\lambda_{\hat{k}|0}^{0},\dots],
\end{split}
\end{equation}
where $\lambda\in \mathbb{R}^{|\mathcal{SS}^0|}$ and $\mathcal{J}^{0}_{\text{wc},\hat{k}|0}=0$ for $\hat{k}\geq \bar{N}$. 
Furthermore, we define the worst-case cost-to-go for each trajectory in the set~\eqref{eq: sample set} recursively as
\begin{equation}
    \begin{split}
        \mathcal{J}^{\hat{h}}_{\text{wc},N|\hat{t}} =&  Q^{\hat{h}-1}(\lambda^{\hat{h}}_{\hat{t}}),\\
        \mathcal{J}^{\hat{h}}_{\text{wc},\hat{k}|\hat{t}} =& \ell_{\text{max}}(x^{\hat{h}}_{\hat{k}|\hat{t}},v^{\hat{h}}_{\hat{k}|\hat{t}},s^{\hat{h}}_{\hat{k}|\hat{t}})-\ell_{\text{max,s}} + \mathcal{J}^{\hat{h}}_{\text{wc},\hat{k}+1|\hat{t}},
    \end{split}
\label{eq: worst case cost to go}
\end{equation}
for $\hat{k}=1,\dots,N-1$. 
The terminal cost is the convex combination of the worst-case cost-to-go:
\begin{equation}
\label{eq: terminal cost}
    \begin{split}      Q^h(\lambda)&= \sum_{\hat{h}=0}^{h} \sum_{\hat{t}=0}^{\infty} \sum_{\hat{k}=0}^{N-1} \lambda_{\hat{k}|\hat{t}}^{\hat{h}} \mathcal{J}^{\hat{h}}_{\text{wc},\hat{k}|\hat{t}},\\
        \text{s.t} & \quad \lambda=[\dots,\lambda_{\hat{k}|\hat{t}}^{\hat{h}},\dots]. \\           
   \end{split}
\end{equation}
\begin{remark}
Given $Q(\lambda)$ and $\mathcal{CS}^h_{\text{robust}}$, we can define the optimal cost-to-go as $Q^{h,\star}(x,s)=\min_{(x,s,\lambda)\in\mathcal{CS}^h_{\text{robust}}}Q(\lambda)$. This cost exhibits a more direct dependence on the terminal state  $(x_{N|t}^h, s_{N|t}^h)$, similar to the notation in~\cite{rosolia2017learning}. However, this notation would make some of the technical arguments in the subsequent analysis more cumbersome, which is why we use both the cost $Q^h$ and the set $\mathcal{CS}^h_{\text{robust}}$ separately in the analysis. Although we define the sample set $\mathcal{SS}^h$ in~\eqref{eq: sample set} using infinitely long closed-loop trajectories with $t\rightarrow\infty$, practical application typically only uses trajectories of finite time~\cite{rosolia2017learning}.
\end{remark}

\subsection{Offline and Online Algorithms}\label{Algorithms}
The computation is divided into an offline part and an online part. Wherever possible, computationally intensive tasks are delegated to the offline phase, as described in Algorithm~\ref{alg:Offline}. During the online phase, outlined in Algorithm~\ref{alg:Online}, the convex quadratic program~\eqref{eq:RALMPC} is solved at each time step $t$. The resulting data from the online execution is then used to refine $\Theta^{\text{HC},h}$ and expand the set $\mathcal{SS}^h$. More discussion on the computational complexity is provided in 
\ifthenelse{\boolean{arxivversion}}{Appendix~\ref{Computational Complexity}}{~\cite{petrenz2025robust}}.
\begin{algorithm}
\caption{\textbf{Robust Adaptive Learning MPC - Offline}}
\label{alg:Offline}
\begin{algorithmic}
\Statex Compute feedback K (Assumption~\ref{ass: controller}).
\Statex Design the polytope $\mathcal{P}$~\eqref{eq: polytope}.
\Statex Compute $\rho_{\bar{\theta}_0}$, $L_{\mathbb{B}}$, $\bar{d}$, and $c_j$~\eqref{eq:contraction rate},~\eqref{eq: lipschitz},~\eqref{eq: disturbance bound},~\eqref{eq: constraint bound}.
\Statex Compute an initial solution (Assumption~\ref{ass:initial_solution}) and construct $\mathcal{CS}_{\text{robust}}^{0}$ and $Q^{0}$~\eqref{eq: convex sample safe set},~\eqref{eq:initial terminal cost}.
\end{algorithmic}
\end{algorithm}
\vspace{-0.5cm} 
\begin{algorithm}
\caption{\textbf{Robust Adaptive Learning MPC - Online}}
\label{alg:Online}
\begin{algorithmic}
\Statex \textbf{For} each iteration h, starting at  $x_0^h=x_s$, $\Theta^{\text{HC},h-1}$ \textbf{do}:
\Statex \hspace{0.5cm} \textbf{For} each time $t$, given $x_t^h$ \textbf{do}:
\Statex \hspace{1cm} Update $\Theta_t^{\text{HC},h}$ (Alg.~\ref{alg:hypercube}), $\rho_{\bar{\theta}_t^h}$~\eqref{eq:contraction rate}
\Statex \hspace{1cm} Solve MPC optimization problem~\eqref{eq:RALMPC}
\Statex \hspace{1cm} Apply control input $u_t^h=v_{0|t}^{h,*}+Kx_t^h$
\Statex \hspace{0.5cm} Update terminal condition~\eqref{eq: convex sample safe set},~\eqref{eq: terminal cost}
\end{algorithmic}
\end{algorithm}

\section{Theoretical Analysis}\label{Theoretical Analysis}
In this section, we show that the proposed MPC scheme is recursively feasible, ensures constraint satisfaction, and achieves a desirable closed-loop performance bound.
First, we ensure robust control invariance of the learned terminal set $\mathcal{CS}^h_{\text{robust}}$:
\begin{proposition}
\label{prop: RCI Set}
Suppose Assumptions~\ref{ass: general},~\ref{ass: controller},~\ref{ass: contractive}, and~\ref{ass:initial_solution} hold.  
Then, the set $\mathcal{CS}^h_{\text{robust}}$, as defined in~\eqref{eq: convex sample safe set}, is a robust control invariant set subject to the constraints~\eqref{eq: constraints}, i.e.,  
for all $(x, s,\lambda) \in \mathcal{CS}^h_{\text{robust}}$, $h\in\mathbb{N}$ and any $(\bar{\theta},\eta)$ with
$\bar{\theta}\oplus\eta \mathbb{B}_p\subseteq \Theta_0^{\text{HC}}$, there exists a control input $v\in \mathbb{R}^m$ and $\lambda^+\in \mathbb{R}^{|\mathcal{SS}^h|}$ such that  
\begin{align}
    \big(x^+, s^+ ,\lambda^+\big) &\in \mathcal{CS}^{h}_{\text{robust}},  \\
    \big(x , Kx+v \big) &\in \mathcal{Z},\\
        \label{eq: decrease terminal cost +}
        Q^h(\lambda^+)-Q^h(\lambda)&\leq -\ell_{\mathrm{max}}(x,v,s)+\ell_{\mathrm{max,s}},
\end{align}  
where $x^+ = A_{\text{cl},\bar{\theta}} x + B_{\bar{\theta}} v$ and $s^+ = (\rho_{\bar{\theta}} + \eta L_{\mathbb{B}}) s + \bar{d} + w_{\eta}(x, Kx + v)$.
\end{proposition}
\proof
\ifthenelse{\boolean{arxivversion}}{The proof is provided in Appendix~\ref{Proof of Proposition 1}.
}
{The proof is provided in~\cite{petrenz2025robust}.}
\endproof
The following theorem utilizes the properties of the learned terminal set and cost (Prop.~\ref{prop: RCI Set}) to establish the closed-loop properties of the proposed MPC scheme.
\begin{theorem}
\label{theorem}
    Suppose Assumption~\ref{ass: general},~\ref{ass: controller},~\ref{ass: contractive}, and~\ref{ass:initial_solution} hold.
    Then Problem~\ref{eq:RALMPC} is feasible for all $h\in\mathbb{N}$, $t\in\mathbb{N}$ for the closed-loop system with $u_t^h=v_{0|t}^{h,*}+Kx_t^h$ and  the constraints~\eqref{eq: constraints} are satisfied. 
    Moreover, the closed-loop cost satisfies
    \begin{equation}
     \label{eq:performance_bound}\limsup_{T\rightarrow\infty}\dfrac{1}{T}\sum_{t=0}^{T-1} \ell(x_t^h,u_t^h)\leq \ell_{\mathrm{max,s}}.
    \end{equation}
\end{theorem}

\proof
\ifthenelse{\boolean{arxivversion}}{
The proof is provided in Appendix~\ref{Proof of Theorem 1}.
    }{The proof is provided in~\cite{petrenz2025robust}.}
\endproof
The proof derives a decrease on the optimal cost, which mirrors the Lyapunov condition to show practical asymptotic stability of the closed-loop system~\cite[Th. 2.20]{grune2017nonlinear}.
A formal proof of practical asymptotic stability is given in~\cite[Thm.~4.14]{PetrenzMasterThesis}
\section{Simulation Results}\label{Simulation Results}
The following example illustrates the performance and computational efficiency improvements of the proposed robust adaptive learning MPC scheme compared to robust adaptive MPC (RAMPC)~\cite{kohler2019linear} for an iterative task.
We consider a mass-spring-damper system:
\begin{equation}
    m \ddot{y} = -c \dot{y} - k y + u + d,
\end{equation}
with a fixed mass $m = 1$, an uncertain damping coefficient $c \in [0.1, 0.3]$, and a spring constant $k \in [0.5, 1.5]$. Additionally, the system is affected by an additive disturbance $|d_t| \leq 0.2$. The unknown true values are $c^* = 0.3$ and $k^* = 0.5$. Moreover, the moving window $M$ is set to $10$ for all experiments. We use Euler discretization with a sampling time of $T_s = 0.1$.
Transforming the second-order ODE into a state-space model yields the state $x = (y, \dot{y}) \in \mathbb{R}^2$. The constraints are given by $\mathcal{Z} = [-0.2, 4.1] \times [-5, 5] \times [-15, 15]$.
We consider the control goal to iteratively steer the system from $x_s = \begin{bmatrix}4 & 0\end{bmatrix}^\top$ to the origin, while minimizing the quadratic stage cost with $Q = \mathrm{diag}(1, 10^{-2})$ and $R = 10^{-1}$.
The controller $K$ is computed using linear matrix inequalities, as done in~\cite{kohler2019linear}. Furthermore, the polytope $\mathcal{P}$ is the maximal $\rho$-contractive set under the constraints.
The initial solution in Assumption~\ref{ass:initial_solution} for the convex safe set~\eqref{eq: convex sample safe set} is computed by solving Problem~\eqref{eq:RALMPC} with a finite horizon and a robust positively invariant terminal set, based on~\cite[Proposition 3]{kohler2019linear}. Within the terminal set, we append steps using the control law $Kx$ until reaching the steady state defined in Assumption~\ref{ass:initial_solution}.

First, we investigate how changing the horizon $N$ affects the computation time and resulting cost in comparison to RAMPC. To obtain comparable results over the iterations, we set the disturbances $d_t$ to a constant value, $d_t = 0.1$.
Both algorithms are repeated over 20 iterations, where the set $\Theta^{\text{HC}}$ is further reduced from iteration to iteration. We solve the problem using CasADi~\cite{Andersson2019} and IPOPT, and the code is publicly available online: {\footnotesize\url{https://github.com/HannesPetrenz/RALMPC_Linear_Uncertain_Systems}}.  
The RAMPC algorithm requires a horizon of at least $N=22$ steps to be feasible, so we set the horizon to $N_{\text{RAMPC}}=25$. 
Table~\ref{tab:comparison} shows an overall improvement in the cost of the final iteration for horizon $N_{\text{RALMPC}}=\{8,12,18\}$ and average CPU time compared to RAMPC. The improvement in computation time is due to the reduced horizon, which has a larger impact than the additional optimization variable $\lambda$. Note that the dimension of $\lambda$ grows with the size of the sample set, emphasizing the importance of reducing the sample set to only promising states. 
Table~\ref{tab:comparison} shows the trade-off between cost improvement and computation time increase in comparison to the optimal solution (OS), which solves the robust infinite horizon optimization problem with knowledge of the true parameters, accounting only robustly for the unknown disturbances $d_t$
\begin{table}[t]
    \centering 
    \renewcommand{\arraystretch}{1.1} 
    \setlength{\tabcolsep}{3pt} 
    \begin{tabular}{c|cc|ccc}
        \toprule
        \multirow{2}{*}{\textbf{N}} & \multicolumn{2}{c|}{\textbf{Avg. Comp. Time (s)}} & \multicolumn{3}{c}{\textbf{Cost}} \\
        \cline{2-6}
        & RAMPC & RALMPC &OS & RAMPC & RALMPC \\
        \midrule
        25  & 11.9 & -   & - & 159 & -    \\
        18  & -   & 8.0  & - & -   & 139  \\
        12  & -   & 3.6  & - & -   & 140  \\
        8   & -   & 1.8  & - & -   & 149  \\
        6   & -   & 1.3  & - & -   & 161  \\
        $\infty$   & -   & -    &135& -   & -  \\
        \bottomrule
    \end{tabular}
    \caption{Comparison of RAMPC and RALMPC in terms of average computation time solving the MPC problem and closed-loop cost of the last iteration.}
    \label{tab:comparison}
\vspace{-5mm}
\end{table}

\begin{figure}
    \centering
    \includegraphics[width=0.4\textwidth]{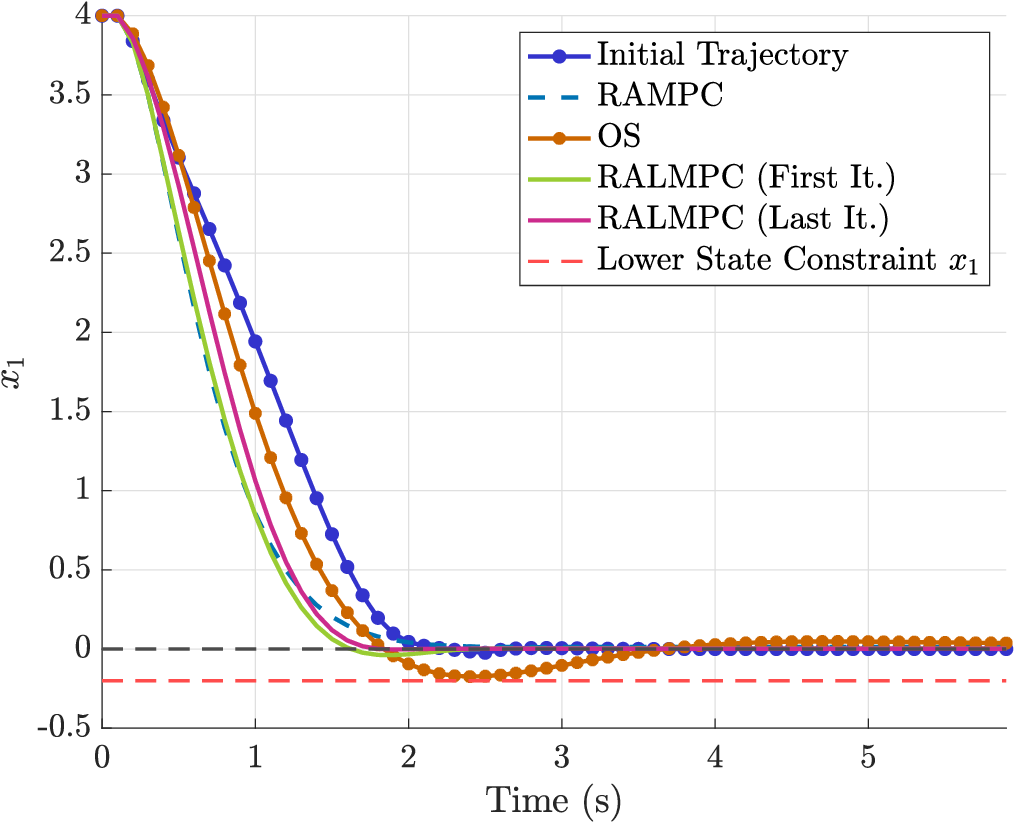}
    \caption{Comparison of RALMPC ($N_{\text{RALMPC}}=12$) and RAMPC trajectories for state $x_1$
    over time. The initial trajectory (blue) is contained in the initial sample set. The optimal solution (OS, orange) represents the theoretical best achievable performance, which serves as a benchmark. RALMPC (green: first iteration, magenta: last iteration) illustrates the iterative improvement of RALMPC.}
    \label{fig:PaperPlot_timex1} 
\vspace{-5mm}
\end{figure}

Figure~\ref{fig:PaperPlot_timex1} shows the closed-loop trajectories for the different controllers. The horizon length of RALMPC is set to $N_{\text{RALMPC}} = 12$, as it provides the best trade-off between computation time and cost. It clearly demonstrates that the RALMPC algorithm approaches the optimal solution through learning. Moreover, it reveals that, even though the horizon $N_{\text{RAMPC}}$ is twice as long, the trajectory of RAMPC remains far from the optimal trajectory, demonstrating the advantage of RALMPC.
\section{Conclusion}\label{Conclusion}
We proposed a robust adaptive learning MPC framework for linear systems with parametric uncertainties and additive disturbances. We extended the robust adaptive MPC in~\cite{kohler2019linear} to iterative tasks by learning the terminal cost and terminal set. The proposed algorithm is recursively feasible, ensures robust constraint satisfaction, and guarantees a robust bound on the closed-loop cost. In a numerical example, we demonstrate the advantages of the method for iterative tasks compared to~\cite{kohler2019linear}. The RALMPC decreases the closed-loop cost function by iterative learning. Moreover, it allows for reducing the horizon of the MPC, which leads to a reduced computational complexity. 
An open issue is improving computational efficiency by suitably selecting trajectories in the sample set.
\ifthenelse{\boolean{arxivversion}}{

\appendices
\section{}
In this appendix, we first analyze the computational complexity of the proposed algorithm in comparison with existing approaches.
We then provide the proofs of Proposition~\ref{prop: RCI Set} and Theorem~\ref{theorem}.
\subsection{Computational Complexity} \label{Computational Complexity}
We now analyze the computational complexity of the proposed algorithm, particularly in comparison with the RAMPC in~\cite{kohler2019linear}, see Table~\ref{tab: Computational Complexity}. The evaluation of $w_{\eta}(x,u)$ and the terminal constraint introduces $N \cdot r \cdot 2^p$ and $r + |\mathcal{SS}^h| + 1$ additional inequality constraints, respectively, compared with a nominal MPC. Moreover, the algorithm requires $N + 2 + |\mathcal{SS}^h|$ additional decision variables due to the introduction of $s$ and $\lambda$. In contrast, the RAMPC in~\cite{kohler2019linear} only requires $N \cdot r \cdot 2^p$ additional inequality constraints and $N + 1$ additional variables.

\begin{table}[h]
\centering
\begin{tabular}{|c|c|c|}
\hline
 & RAMPC & RALMPC \\
\hline
Ineq. constraints & $N \cdot r \cdot 2^p$ & $N \cdot r \cdot 2^p + r + |\mathcal{SS}^h| + 1$ \\
\hline
Decision variables & $N + 1$ & $N + |\mathcal{SS}^h| + 2$ \\
\hline
\end{tabular}
\caption{Comparison of the additional computational complexity introduced by RAMPC and the proposed RALMPC relative to a nominal MPC.}
\label{tab: Computational Complexity}
\end{table}

\subsection{Proof of Proposition 1} \label{Proof of Proposition 1}
In the following, we provide the proof of Proposition~\ref{prop: RCI Set}, establishing the control-invariant set property, and then prove Equation~\ref{eq: decrease terminal cost +} in Part~2.
\begin{proof}
\textbf{Part 1:}
    Given that $(x,s,\lambda) \in \mathcal{CS}^h_{\text{robust}}$, there exists $\lambda=[\dots, \lambda_{\hat{k}|\hat{t}}^{\hat{h}}, \dots]$ with $\sum_{\hat{h}=0}^{h} \sum_{\hat{t}=0}^{\infty} \sum_{\hat{k}=0}^{N-1} \lambda_{\hat{k}|\hat{t}}^{\hat{h}}=1$ and $\lambda_{\hat{k}|\hat{t}}^{\hat{h}}\geq0$ such that $(x',v',s')=\sum_{\hat{h}=0}^{h} \sum_{\hat{t}=0}^{\infty} \sum_{\hat{k}=0}^{N-1} \lambda_{\hat{k}|\hat{t}}^{\hat{h}} (x_{\hat{k}|\hat{t}}^{\hat{h}},v_{\hat{k}|\hat{t}}^{\hat{h}},s_{\hat{k}|\hat{t}}^{\hat{h}})$ satisfies $\underset{i}{\max} H_i(x-x')\leq s'-s$. 
    We denote $({\bar{t}},{\bar{h}})=\arg\min_{\hat{t},\hat{h}} \eta_{\hat{t}}^{\hat{h}}$. 
   Recall that $(x_{N|\hat{t}}^{\hat{h}},s_{N|\hat{t}}^{\hat{h}},\lambda_{\hat{t}}^{\hat{h}})\in\mathcal{CS}^{\hat{h}-1}_{\text{robust}}$, i.e., there exists a $(\tilde{x}_{\hat{t}}^{\hat{h}},\tilde{s}_{\hat{t}}^{\hat{h}},\lambda_{\hat{t}}^{\hat{h}})\in\mathcal{CS}^{\hat{h}-1}$ such that $\max_iH_i(x_{N|\hat{t}}^{\hat{h}}-\tilde{x}_{\hat{t}}^{\hat{h}})\leq\tilde{s}_{\hat{t}}^{\hat{h}}-s_{N|\hat{t}}^{\hat{h}}$ for all $\hat{h}<h$.
    Let us define
    \begin{align}
        \label{eq: convex dynamic} \notag
            &x'^{+}:=\sum_{\hat{h}=0}^{h} \sum_{\hat{t}=0}^{\infty} \sum_{\hat{k}=0}^{N-2} \lambda_{\hat{k}|\hat{t}}^{\hat{h}} x_{\hat{k}+1|\hat{t}}^{\hat{h}}+\sum_{\hat{h}=0}^{h} \sum_{\hat{t}=0}^{\infty} \lambda_{N-1|\hat{t}}^{\hat{h}} \tilde{x}_{\hat{t}}^{\hat{h}}\\ \notag
            =&\sum_{\hat{h}=0}^{h} \sum_{\hat{t}=0}^{\infty} \sum_{\hat{k}=0}^{N-1} \lambda_{\hat{k}|\hat{t}}^{\hat{h}} x_{\hat{k}+1|\hat{t}}^{\hat{h}}-\sum_{\hat{h}=0}^{h} \sum_{\hat{t}=0}^{\infty} \lambda_{N-1|\hat{t}}^{\hat{h}} (x_{N|\hat{t}}^{\hat{h}}-\tilde{x}_{\hat{t}}^{\hat{h}})\\ \notag
            \stackrel{~\eqref{eq:RALMPC b}}{=}& A_{\text{cl},\bar{\theta}} x' + B_{\bar{\theta}} v' - D(x',u') \Delta \bar{\theta}_{\bar{t}}^{\bar{h}} \\ 
            &-\sum_{\hat{h}=0}^{h} \sum_{\hat{t}=0}^{\infty} \left(  \sum_{\hat{k}=0}^{N-1} \lambda_{\hat{k}|\hat{t}}^{\hat{h}} D(x_{\hat{k}|\hat{t}}^{\hat{h}},u_{\hat{k}|\hat{t}}^{\hat{h}}) \Delta\bar{\theta}_{\hat{t}}^{\hat{h}}\right. \\ \notag
            &\left. +\lambda_{N-1|\hat{t}}^{\hat{h}} (x_{N|\hat{t}}^{\hat{h}}-\tilde{x}_{\hat{t}}^{\hat{h}})\right), \notag
    \end{align}
    where $\Delta \bar{\theta}_{\bar{t}}^{\bar{h}} = \bar{\theta} - \bar{\theta}_{\bar{t}}^{\bar{h}}$, $\Delta \bar{\theta}_{\hat{t}}^{\hat{h}} = \bar{\theta}_{\bar{t}}^{\bar{h}} - \bar{\theta}_{\hat{t}}^{\hat{h}}$ and $u'=Kx'+v'$.
    
    Since it holds that $\theta_{\bar{t}}^{\bar{h}}-\theta_{\hat{t}}^{\hat{h}} \in \Delta\eta_{\hat{t}}^{\hat{h}} \mathbb{B}_p$ with $\Delta\eta_{\hat{t}}^{\hat{h}}=\eta_{\hat{t}}^{\hat{h}} - \eta_{\bar{t}}^{\bar{h}}$, we can lower bound  
    \begin{align}
         \label{eq: convex tube dynamic}
        &s'^+\\ \ \notag
        :=&\sum_{\hat{h}=0}^{h} \sum_{\hat{t}=0}^{\infty} \sum_{\hat{k}=0}^{N-2} \lambda_{\hat{k}|\hat{t}}^{\hat{h}} s_{\hat{k}+1|\hat{t}}^{\hat{h}}+\sum_{\hat{h}=0}^{h} \sum_{\hat{t}=0}^{\infty} \lambda_{N-1|\hat{t}}^{\hat{h}} \tilde{s}_{\hat{t}}^{\hat{h}}\\ \notag
        =&\sum_{\hat{h}=0}^{h} \sum_{\hat{t}=0}^{\infty} \sum_{\hat{k}=0}^{N-1} \lambda_{\hat{k}|\hat{t}}^{\hat{h}} s_{\hat{k}+1|\hat{t}}^{\hat{h}}-\sum_{\hat{h}=0}^{h} \sum_{\hat{t}=0}^{\infty} \lambda_{N-1|\hat{t}}^{\hat{h}} (s_{N|\hat{t}}^{\hat{h}}-\tilde{s}_{\hat{t}}^{\hat{h}})\\ \notag
        =&\sum_{\hat{h}=0}^{h} \sum_{\hat{t}=0}^{\infty} \sum_{\hat{k}=0}^{N-1} \lambda_{\hat{k}|\hat{t}}^{\hat{h}} \left[(\rho_{\bar{\theta}_{\hat{t}}^{\hat{h}}} +\eta_{\hat{k}|\hat{t}}^{\hat{h}} L_{\mathbb{B}})s + \bar{d}  + w_{\eta}(x_{\hat{k}|\hat{t}}^{\hat{h}}, u_{\hat{k}|\hat{t}}^{\hat{h}})\right]\\ \notag
        &-\sum_{\hat{h}=0}^{h} \sum_{\hat{t}=0}^{\infty} \lambda_{N-1|\hat{t}}^{\hat{h}} (s_{N|\hat{t}}^{\hat{h}}-\tilde{s}_{\hat{t}}^{\hat{h}})\\ \notag
        \geq&\rho_{\bar{\theta}_{\bar{t}}^{\bar{h}}} s'+w'+\sum_{\hat{h}=0}^{h} \sum_{\hat{t}=0}^{\infty} \sum_{\hat{k}=0}^{N-1} \lambda_{\hat{k}|\hat{t}}^{\hat{h}} w_{\Delta\eta_{\hat{t}}^{\hat{h}}}(x_{\hat{k}|\hat{t}}^{\hat{h}},u_{\hat{k}|\hat{t}}^{\hat{h}})\\ \notag
        &+\sum_{\hat{h}=0}^{h} \sum_{\hat{t}=0}^{\infty} \lambda_{N-1|\hat{t}}^{\hat{h}} (\tilde{s}_{\hat{t}}^{\hat{h}}-s_{N|\hat{t}}^{\hat{h}}),\notag
    \end{align}
    by using the inequality $\rho_{\theta_{\bar{t}}^{\bar{h}}}-\Delta\eta_{\hat{t}}^{\hat{h}}L_{\mathbb{B}}\leq\rho_{\theta_{\hat{t}}^{\hat{h}}}$~\cite[Prop. 1]{kohler2019linear} and $w'=\bar{d}+\eta_{\bar{t}}^{\bar{h}} L_{\mathbb{B}} s'+w_{\eta_{\bar{t}}^{\bar{h}}}(x',u')$.
    Note that $(\tilde{x}_{\hat{t}}^{\hat{h}},\tilde{s}^{\hat{h}}_{\hat{t}},\tilde{\lambda}^{\hat{h}}_{\hat{t}})\in\mathcal{CS}^{\hat{h}-1}$ and $(x_{\hat{k}+1|\hat{t}}^{\hat{h}},s_{\hat{k}+1|\hat{t}}^{\hat{h}})\in \mathcal{SS}^{h}$, $k=0,\dots,N-2$ ensures that there exists a $\lambda^+$ such that $(x^{'+},s^{'+},\lambda^+)\in\mathcal{CS}^h$.
    Additionally, we define the auxiliary tube dynamic 
    \begin{equation}
    \label{eq: convex auxilary tube dynamic}
        \begin{split}
        \tilde{s}'^+=&\rho_{\bar{\theta}}\tilde{s}'+w_{\Delta \eta_{\bar{t}}^{\bar{h}}}(x',u')\\
        &+\sum_{\hat{h}=0}^{h} \sum_{\hat{t}=0}^{\infty} \sum_{\hat{k}=0}^{N-1} \lambda_{\hat{k}|\hat{t}}^{\hat{h}} w_{\Delta \eta_{\hat{t}}^{\hat{h}}}(x_{\hat{k}|\hat{t}}^{\hat{h}},u_{\hat{k}|\hat{t}}^{\hat{h}})\\
        &+\sum_{\hat{h}=0}^{h} \sum_{\hat{t}=0}^{\infty} \lambda_{N-1|\hat{t}}^{\hat{h}} (\tilde{s}_{\hat{t}}^{\hat{h}}-s_{N|\hat{t}}^{\hat{h}}),
        \end{split}
    \end{equation}
    with $\tilde{s}'=s'-s$.
   Next, we show that the auxiliary tube $\tilde{s}'\cdot\mathcal{P}$ bounds the error between $x'^+$ and $x^+=A_{\text{cl},\bar{\theta}} x + B_{\bar{\theta}} v'$, i.e., $x^+-x'^+=e^+\in \tilde{s}'^+\cdot\mathcal{P}$. Using $\underset{i}{\max} H_i(x-x')\leq s'-s=\tilde{s}'$, it holds that
    \begin{equation}
    \label{eq: bounded error}
    \begin{split}
        &\underset{i}{\max}H_i(x^+-x'^+)\\
        \stackrel{~\eqref{eq: convex dynamic}}{=}&\underset{i}{\max} H_i\left(A_{\text{cl},\bar{\theta}}(x-x')+ D(x',u') \Delta \bar{\theta} _{\bar{t}}^{\bar{h}}\right.\\
        &+\sum_{\hat{h}=0}^{h} \sum_{\hat{t}=0}^{\infty} \sum_{\hat{k}=0}^{N-1} \lambda_{\hat{k}|\hat{t}}^{\hat{h}} D(x_{\hat{k}|\hat{t}}^{\hat{h}},u_{\hat{k}|\hat{t}}^{\hat{h}}) \Delta \bar{\theta}_{\hat{t}}^{\hat{h}}\\
         &\left. +\sum_{\hat{h}=0}^{h} \sum_{\hat{t}=0}^{\infty} \lambda_{N-1|\hat{t}}^{\hat{h}} (x_{N|\hat{t}}^{\hat{h}}-\tilde{x}_{\hat{t}}^{\hat{h}})\right)\\
        \leq&\rho_{\bar{\theta}}\tilde{s}'+w_{\Delta \eta_{\bar{t}}^{\bar{h}}}(x',u')\\
        &+\sum_{\hat{h}=0}^{h} \sum_{\hat{t}=0}^{\infty} \sum_{\hat{k}=0}^{N-1} \lambda_{\hat{k}|\hat{t}}^{\hat{h}} w_{\Delta \eta_{\hat{t}}^{\hat{h}}}(x_{\hat{k}|\hat{t}}^{\hat{h}},u_{\hat{k}|\hat{t}}^{\hat{h}})\\
        &+\sum_{\hat{h}=0}^{h} \sum_{\hat{t}=0}^{\infty} \lambda_{N-1|\hat{t}}^{\hat{h}} (\tilde{s}_{\hat{t}}^{\hat{h}}-s_{N|\hat{t}}^{\hat{h}})=s'^+,
    \end{split}
    \end{equation}
    where the inequality uses~\eqref{eq:contraction rate},~\eqref{eq: parametric uncertainty bound},~\eqref{eq: convex dynamic},~\eqref{eq: convex tube dynamic}.
    Thus, $(x^+,s^+,\lambda^+)\in \mathcal{CS}^h_{\text{robust}}$ reduces to $s^++\tilde{s}'^+-s'^+\leq0$ with $s^+=\rho_{\bar{\theta}} s + w$ and $w=\eta L_{\mathbb{B}}s+\bar{d}+w_{\eta}(x, Kx+v)$. Similar to proof of~\cite[Th. 1]{kohler2019linear}, this can be shown by using~\eqref{eq:RALMPC d},~\eqref{eq: convex tube dynamic} and~\eqref{eq: convex auxilary tube dynamic}. Hence, we showed that  $(x^+,s^+,\lambda^+) \in \mathcal{CS}^h_{\text{robust}}$ for $\lambda^+$ and $v=v'$. 
    Finally, the constraint satisfaction~\eqref{eq: constraints} follows with
    \begin{equation}
    \begin{split}
        &F_j x + G_j u + c_j s\leq F_j x' + G_j u'+c_j s'\\
        &=\sum_{\hat{h}=0}^{h} \sum_{\hat{t}=0}^{\infty} \sum_{\hat{k}=0}^{N-1} \lambda_{\hat{k}|\hat{t}}^{\hat{h}}(F_j x_{\hat{k}|\hat{t}}^{\hat{h}} + G_j u_{\hat{k}|\hat{t}}^{\hat{h}} + c_j s_{\hat{k}|\hat{t}}^{\hat{h}} )\leq 0,
    \end{split}
    \end{equation}
    using the fact that all data points satisfy the tightened constraints~\eqref{eq:RALMPC f}. \\
\textbf{Part 2:}    
Using Definition~\eqref{eq: terminal cost}, we get
\begin{equation}
    \begin{split}
        &Q^h(\lambda^+)-Q^h(\lambda)\\
        =& \sum_{\hat{h}=0}^{h} \sum_{\hat{t}=0}^{\infty} \sum_{\hat{k}=0}^{N-1} \lambda_{\hat{k}|\hat{t}}^{\hat{h}} \big( \mathcal{J}^{\hat{h}}_{\text{wc},k+1|\hat{t}} \\
        &- \left(\ell_{\text{max}}(x_{\hat{k}|\hat{t}}^{\hat{h}},v_{\hat{k}|\hat{t}}^{\hat{h}},s_{\hat{k}|\hat{t}}^{\hat{h}})-\ell_{\text{max,s}}\right)-\mathcal{J}^{\hat{h}}_{\text{wc},k+1|\hat{t}} \big) \\
        =&\sum_{\hat{h}=0}^{h} \sum_{\hat{t}=0}^{\infty} \sum_{\hat{k}=0}^{N-1} \lambda_{\hat{k}|\hat{t}}^{\hat{h}}(-\ell_{\text{max}}(x_{\hat{k}|\hat{t}}^{\hat{h}},v_{\hat{k}|\hat{t}}^{\hat{h}},s_{\hat{k}|\hat{t}}^{\hat{h}})+\ell_{\text{max,s}})\\
        \leq&-\ell_{\text{max}}(x',v,s')+\ell_{\text{max,s}},
    \end{split}
\end{equation}
where $\ell_{\text{max}}(x',v,s')$ is a lower bound on the convex combination.
Moreover, using \eqref{eq: decrease stage cost}, we obtain $-\ell_{\text{max}}(x',v,s')\leq-\ell_{\text{max}}(x,v,s)$ for all$(x,s)\in \mathbb{R}^{n+1}$ with $x \oplus s \mathcal{P} \subseteq  x' \oplus s' \mathcal{P}=\mathbb{X}'$.
\end{proof}
\subsection{Proof of Theorem 1}\label{Proof of Theorem 1}
In the following, we provide the proof of Theorem~\ref{theorem}, consisting of two parts: Part~1 establishes recursive feasibility, and Part~2 shows the cost decrease property.
\begin{proof}
    \textbf{Part 1:}
    Recursive feasibility is proved by induction. At $t=0$, the initial solution (Ass.~\ref{ass:initial_solution}) is a feasible solution to the problem~\eqref{eq:RALMPC}. 
    At time $t+1$, given a feasible solution to~\eqref{eq:RALMPC} at $t$, we consider the candidate inputs $v_{k|t+1}^h=v_{k+1|t}^{h,*}$ and $v_{N-1|t+1}^h=v$ according to Proposition \ref{prop: RCI Set}. 
    Furthermore, we define $x_{N+1|t}^{h,*}$, $s_{N+1|t}^{h,*}$ with \eqref{eq:RALMPC b} and \eqref{eq:RALMPC d} using $v_{N|t}^*=v_{N-1|t+1}^h$.
    The error between the candidate solution $x_{k|t+1}^h$ and the previous optimal solution ${x}_{k+1|t}^{h,*}$ is described by the error dynamics
    \begin{equation}
        \label{eq: error dynamic}
        e_{k+1|t+1}^h = A_{\text{cl},\bar{\theta}_{t+1}^h} e_{k|t+1}^h + D(x_{k+1|t}^{h,*}, u_{k+1|t}^{h,*}) \Delta \bar{\theta}_t^h,       
    \end{equation}
    with $\Delta \bar{\theta}_t^h = \bar{\theta}_{t+1}^h - \bar{\theta}_t^h$ and $e_{0|t+1}^h=x^h_{t+1}-x_{1|t}^{h,*}$.
    Additionally, we define the dynamic of the auxiliary tube
    \begin{equation}
    \label{eq: auxiliary tube dynamic}
                \tilde{s}_{k+1|t+1}^h = \rho_{\bar{\theta}_{t+1}^h} \tilde{s}_{k|t+1}^h + w_{\Delta \eta_t^h} (x_{k+1|t}^{h,*}, u_{k+1|t}^{h,*}),
    \end{equation}
    with $\tilde{s}_{0|t+1}^h=s_{0|t+1}^h-s_{1|t}^{h,*}=w_{0|t}^{h,*}$ and $\Delta \eta_t^h= \eta_t^h-\eta_{t+1}^h$. 
    The auxiliary tube dynamic is used to bound the error dynamic~\eqref{eq: error dynamic} between the candidate solution and the previous optimal solution, leading to the condition $x_{k|t+1}^h-x_{k+1|t}^{h,*}=:e_{k|t+1}^h \in \tilde{s}_{k|t+1}^h \cdot \mathcal{P}$. This can be shown similarly to~\eqref{eq: bounded error}.
    
    Next, we prove that the tube of the previous optimal solution contains the tube of the candidate solution, which results in the condition
    \begin{equation}
        s_{k+1|t+1}^h - s_{k+2|t}^{h,*} + \tilde{s}_{k+1|t+1}^h\leq 0.
    \end{equation}
    Similar to proof of~\cite[Th. 1]{kohler2019linear}, this can be shown by using~\eqref{eq:RALMPC d} and~\eqref{eq: auxiliary tube dynamic}.
    Constraint satisfaction follows similarly to the proof of Proposition~\ref{prop: RCI Set}. 
    The last step is to prove satisfaction of the terminal constraint~\eqref{eq:RALMPC h}. 
    Recall that $\underset{i}{\max} H_i (x_{N|t+1}^h-x_{N+1|t}^{h,*})\leq s_{N+1|t}^{h,*}-s_{N|t+1}^h$. 
    Using Proposition \ref{prop: RCI Set}, $(x_{N|t}^{h,*},s_{N|t}^{h,*},\lambda_{t}^{h,*})\in\mathcal{CS}^{h-1}_{\text{robust}}$ ensures that there exist an input $v$ and a $\lambda^+$ such that
    $(x_{N+1|t}^{h,*},s_{N+1|t}^{h,*},\lambda^+)\in\mathcal{CS}^{h-1}_{\text{robust}}$.
    Furthermore, the definition of $\mathcal{CS}^{h-1}_{\text{robust}} $~\eqref{eq: convex sample safe set} guarantees that there exist $(x'^+,s'^+,\lambda^+)\in \mathcal{CS}^{h-1}$ such that $\underset{i}{\max} H_i (x_{N+1|t}^{h,*} - x'^+) \leq s'^+-s_{N+1|t}^{h,*}$.
    Combining the two inequalities results in 
    \begin{equation}
        \underset{i}{\max} H_i (x_{N|t+1}^h - x'^+)\leq s'^+-s_{N|t+1}^h,\\
    \end{equation}
    which proves that $(x_{N|t+1}^{h},s_{N|t+1}^{h},\lambda^+)\in\mathcal{CS}^h_{\text{robust}}$.
    This completes the recursive feasibility proof. \\
    \textbf{Part 2:} 
    Using the candidate solution at time $t+1$ from the recursive feasibility proof, we derive:
    \begin{equation}
        \begin{split}
            & J^{\text{LMPC},h,*}(x_{t+1}^h,\Theta_{t+1}^{\text{HC},h})-J^{\text{LMPC},h,*}(x_{t}^h,\Theta_t^{\text{HC},h})\\
            \leq& 
            \sum_{k=0}^{N-1} \ell_{\text{max}}(x_{k|t+1}^{h},u_{k|t+1}^{h},s_{k|t+1}^{h}) + Q^{h-1}(\lambda_{t+1}^{h})\\
            &-\sum_{k=0}^{N-1} \ell_{\text{max}}(x_{k|t}^{h,*},u_{k|t}^{h,*},s_{k|t}^{h,*}) - Q^{h-1}(\lambda_{t}^{h,*}) \\
            \leq &-\ell_{\text{max}}(x_{0|t}^{h,*},u_{0|t}^{h,*},s_{0|t}^{h,*})+ Q^{h-1}(\lambda_{t+1}^{h}) \\&+ \ell_{\text{max}}(x_{N-1|t+1}^{h},u_{N-1|t+1}^{h},s_{N-1|t+1}^{h})- Q^{h-1}(\lambda_{t}^{h,*}),
        \end{split}
    \end{equation}
    where the last inequality uses monotonicity~\eqref{eq: decrease stage cost}. 
    From Proposition~\ref{prop: RCI Set}, we further obtain:
    \begin{equation}
    \label{eq: proof cost decay}
        \begin{split}
            & J^{\text{LMPC},h,*}(x_{t+1}^h,\Theta_{t+1}^{\text{HC},h})-J^{\text{LMPC},h,*}(x_{t}^h,\Theta_t^{\text{HC},h}) \\
            \leq&-\ell(x_t^h,u_t^h)+\ell_{\text{max,s}}
        \end{split}
    \end{equation}
    The average stage cost bound~\eqref{eq:performance_bound} directly follows by using the cost decay~\eqref{eq: proof cost decay} in a telescopic sum and the fact that $J^{\mathrm{LMPC},h,*}$ is bounded on the compact feasible set.
\end{proof}
\endproof
}
{}


\bibliographystyle{IEEEtran}
\bibliography{references}

\end{document}